\documentclass[preprint,12pt,authoryear]{elsarticle}
\usepackage{amssymb}
\usepackage{amsmath}
\usepackage[T1]{fontenc}
\usepackage{graphicx}
\usepackage{rotating}
\usepackage{verbatim}
\usepackage{float}
\usepackage[table,xcdraw]{xcolor}
\usepackage[T1]{fontenc}
\usepackage{geometry}
\usepackage[export]{adjustbox}
\usepackage{pdflscape}  
\journal{Computers in Human Behavior}
\begin{document}
\begin{frontmatter}
\title{Topology or Demography? Understanding Sources of Bias in Opinion Dynamics Models}

\author[uni1]{Stanisław Stępień}
\author[uni2]{Michalina Janik}
\author[uni1]{Mateusz Nurek}
\author[uni3]{Akrati Saxena}
\author[uni1]{Radosław Michalski}
\affiliation[uni1]{organization={Wroclaw University of Science and Technology},
             country={Poland}}

\affiliation[uni2]{organization={Aarhus University},
             country={Denmark}}

\affiliation[uni3]{organization={Leiden University},
             country={The Netherlands}}

%COMMENT:
%WHY DO WE USE PREDICTIVE MODELlIING LAYER OF A CLASIFIER INSTEAD OF FOCUSING SOLELY ON THE STATS

%WE utilize logistic regression as well as tree base modelling which is fully interpretable machine leanirng model. We can analyze the decision making process at each every step. 
%We investigate data pattern which go beyond a simple data aggregation, are visible on the unseen data, It learns the pattern visible in the latent space of complex data space. It also allows to seek for general data patterns instead of just aggregated statistics. 

%% Abstract
\begin{abstract}
Ways in which people's opinions change are, without a doubt, subject to a rich tapestry of differing influences. Factors that affect how one arrives at an opinion reflect how they have been shaped by their environment throughout their lives, education, material status, what belief systems they subscribe to, and what socio-economic minorities they are a part of. This already complex system is further expanded by the ever-changing nature of one's social network. It is therefore no surprise that many models have a tendency to perform best for the majority of the population and discriminate against those people who are members of various marginalized groups. This bias and the study of how to counter it are subject to a rapidly developing field of Fairness in Social Network Analysis (SNA). 
\par The focus of this work is to look into how a state-of-the-art model discriminates certain minority groups and whether it is possible to reliably predict for whom it will perform worse. Moreover, is such a prediction possible based solely on one's demographic or topological features? To this end, the NetSense dataset, together with a state-of-the-art CoDiNG model for opinion prediction, has been employed. Our work explores how three classifier models (Demography-Based, Topology-Based, and Hybrid) perform when assessing for whom this algorithm will provide inaccurate predictions. Finally, through a comprehensive analysis of these experimental results, we identify four key patterns of algorithmic bias: 
1. variability in predictive effectiveness across issue types and minority groups, 
2. extreme predictive variability for behaviourally-defined minorities, 
3. potent but variable predictive signals from socio-economic and religious minorities, 
4. persistent challenges in accurately predicting misclassifications for certain demographic minorities. 
Our findings suggest that no single paradigm provides the best results and that there is a real need for context-aware strategies in fairness-oriented social network analysis. We conclude that a multi-faceted approach, incorporating both individual attributes and network structures, is essential for reducing algorithmic bias and promoting inclusive decision-making.
\end{abstract}

%% Keywords
\end{frontmatter}
%%%%%%%%%%%%%%%%%%%%%%%%%%%%%%%%%%%%%%%%%%%%%%%%%%%%%%%%%%%%%%%%%%%%%%%%%
\section{Introduction}
Society can be thought of as a mixture of individuals, each holding and exchanging opinions when interacting. This complex system of opinion dynamics can sometimes start a revolution or, in most cases, simply agree or disagree on something. However, being able to predict people's opinions is an important sociological challenge that is often needed in multiple situations: for instance, when introducing a legislative change in countries or designing urban spaces. Independent of the case, the complexity of this task is high, given that it involves multiple factors, the most important ones being the understanding of cognitive processes, opinion dynamics, and human interactions. This task is assessed in multiple ways, but often researchers overlook an important factor that has recently gained a lot of importance in the research of social network analysis: the aspect of fairness.

In order to train an algorithm to accurately assess which opinion a person is going to have, a multitude of factors should be considered: their choice history, who that person has interacted with in the past, and, obviously, their attributes such as gender, age, or place of living. There is no one, clear-cut recipe on how to create an algorithm that will be able to anticipate future opinions of everyone, and each real-life case requires a deeper insight in order to identify outliers, non-conformists, and other hard-to-guess participants. However, it is generally the case that the more data we have of a certain group of people, the better we are able to train our models to work with such a group. At the foundation of this work lie three questions that we have asked:
\begin{enumerate}
    \item When it comes to opinion prediction, are there any sub-groups for which algorithms perform better or worse?
    \item Can we guess for whom an algorithm will more often mispredict?
    \item Which of the following holds more predictive power in this regard: demographic or topological data of the users?
\end{enumerate}
%Fainess Background
The evolution of real-world social networks is generally affected by user homophily, which is a widespread tendency of people to contact others who are perceived as similar to themselves \cite{BirdsofaFeather}. 
This may be due to personal bias, a product of one's social bubble, or even the unfair influence of an algorithm affecting their choices. This homophilic tendency of a given model does not need to stem from the intentions of algorithm creators. Instead, it often results from the fact that the accuracy of the tools that we have is dependent on the size of the communities for which the model is trained. With this taken into account, one could ask the following question: \textit{Is it one's demographic attributes or rather their place in the network that affects their likelihood of being systematically more misclassified by the model?}
%Motivation
\par
Answering this question and identifying the factors that correlate most strongly with these mispredictions would make it possible to assess how fair or biased a given model is. It would also allow the discovery of any potential vulnerabilities that the given algorithm is prone to.
%Objectives
\par
Therefore, the goal of this work is to construct a transparent framework that would allow such an assessment of the model's misclassification bias and to establish whether it skews more towards demographic or topological factors of the affected people. Moreover, it would be beneficial to distinguish whether the type of survey question for which the opinion is predicted affects how different minorities are affected, or whether, and compare the results with the general population.
%Results
\par
However, the results obtained throughout this work indicate that in the context of predicting which factors are likely to play the most significant role in the model's discriminative behaviour, there is no one answer that is likely to hold true for each type of minority. It appears that consistently both the considered subgroup and the very nature of the question for which an answer is being predicted play a relevant role in predicting when the dip in the opinion classifier's performance is going to occur.

%\subsection{Paper Structure}
\par This work has been structured into eight sections, together with an additional appendix section at the end of the document. The following - Section 2 focuses on the discussion of works within the field of Fair Social Network Analysis, which have been foundational to the creation of this research. Section 3 delves into the details of the CoDiNG model, which served as the basis for the tests conducted. Section 4 focuses on the structure and insights of the NetSense dataset and the definitions for consecutive minorities. It also contains baseline observations derived from the CoDiNG experiment dataset. Section 5 contains the description of our framework's architecture and insights from the process of classifiers' training. Then, Section 6 describes the results of the experiments. Next, Section 7 focuses on the discussion of the obtained results. Finally, Section 8 summarizes the findings of our work, discusses them in the context of fair opinion dynamics, and considers future avenues for this line of research.
\par
The following section will focus on the overview of other works that delve into the problem of fairness and opinion dynamics in the context of SNA, together with a look into the problem surrounding the use of surveys for reporting and the biases it often brings to the obtained data.

\section{Related Work}
\par Although the problem of fairness in the context of social networks is still in its early stages, it has been explored much more extensively when it comes to the field of machine learning. Several frameworks have been proposed that strive to provide fair classification for the individuals who are similar with respect to a particular task~\cite{dwork2012fairness}, prevent potential discrimination in regard to a sensitive attribute in the problem of supervised learning~\cite{hardt2016equality} or to understand the relationship between fairness and algorithmic trade-offs, especially for public safety~\cite{corbett2017algorithmic}. However, these were not designed for the social networks and do not utilize the topological and relational information contained within the structure of a graph.

\subsection{Fairness in Social Network Analysis}
The field of fair social network analysis is much less explored \cite{saxena2024fairsna}, and recently, there have been few works focusing on fairness in different network analysis tasks, such as link prediction \cite{saxena2022nodesim, saxena2022hm}, Influence Maximization \cite{stoica2019fairness}, Influence Blocking \cite{saxena2023fairness}, centrality ranking \cite{tsioutsiouliklis2021fairness} and community detection \cite{de2024group}. It is a rapidly growing area of research with the potential to positively influence the lives of those affected by various online network systems. Many of those personal networks have a tendency to display some form of homophily and therefore affect the way they interact with their environment and what information may reach them~\cite{BirdsofaFeather, saxena2025homophily}. Those structural inequalities tend to disproportionately influence outcomes for different types of minorities within the population and contribute to the general inequality experienced by the under-represented individuals~\cite{karimi2022minorities}. It has been proven that this affects the distribution of the degree ranks among such sub-groups~\cite{karimi2018homophily}.

\subsection{Opinion Dynamics in Social Networks}
Currently, opinion modelling approaches are typically categorized into three groups: discrete, continuous, and hybrid. In discrete models, opinions usually take one of two possible values (e.g., agree or disagree). Binary-state dynamics serve as foundational tools for capturing essential aspects of social interactions. The concept of nodes switching between two states under the influence of their neighbors provides a simplified yet insightful representation of how opinions evolve in a network. Among discrete models, those rooted in statistical physics, such as the Ising model (\cite{ising1925beitrag, glauber1963time, bianconi2002mean}) and the Sznajd model (\cite{sznajd2000opinion}), have gained widespread popularity. Other well-known models include the Voter model (\cite{clifford1973model}) and threshold-based models like the Global Threshold Model (GTM) (\cite{granovetter1978threshold}) and the Network Threshold Model (NTM) \cite{valente1996social}, which model social influence through cumulative pressure from others. The Independent Cascade Model (ICM) (\cite{goldenberg2001talk}) is a probabilistic model focused on pairwise interactions, where each neighbor has a certain probability of triggering an opinion change. Another notable discrete model is the Naming Game (\cite{baronchelli2006sharp}), which introduces a third, ambiguous state—offering a more realistic setting than purely polarized binary models.

In continuous opinion models, opinions are represented as real values, capturing the strength of an agent’s conviction. Popular examples include the classical DeGroot model~(\cite{degroot1974reaching}) and its extension, the Friedkin-Johnsen model~(\cite{friedkin1990social}), which introduces the concept of stubborn agents - individuals who retain their initial opinion. Bounded confidence models rely on the assumption that agents influence each other only if their opinions are sufficiently close. The most popular bounded confidence models are Deffuant–Weisbuch model (DW)~(\cite{deffuant2000mixing}) and the Hegselmann–Krause model (HK)~(\cite{rainer2002opinion}).

Hybrid models combine features of both discrete and continuous approaches. Typically, the continuous component represents latent states, while the discrete part reflects the externally visible opinions used during agent interactions. The family of hybrid models remains relatively small. Among the most prominent are Continuous Opinion and Discrete Action (CODA)~(\cite{martins2008continuous}), Social Network Opinions and Actions Evolutions (SNOAE)~(\cite{zhan2021bounded}), Social Judgment Based Opinion (SJBO)(~\cite{fan2016opinion}), and Continuous-Discrete Naming Game (CoDiNG)~(\cite{CoDING}). Notably, CoDiNG is firmly grounded in cognitive science and has been validated on real-world datasets, demonstrating improved performance over the classical Naming Game model.

Scholars have also examined various phenomena related to opinion dynamics, such as agents' polarization~\cite{amelkin2017polar}, conformity~\cite{das2014modeling}, and the influence of opinion leaders within a population~\cite{zhao2018understanding}. Other studies focus on the role of social network structure in opinion formation~\cite{wu2004socialstructureopinionformation}, as well as on specific factors like network density, which affect how opinions converge within a population~\cite{NguyenDynamicsofopinionformation}.

% Similarly, the study of opinion dynamics in social networks has gained significant attention in recent years, driven by the increasing availability of large-scale datasets and advances in computational methods. Many of the  state-of-the-art models incorporate phenomena such as agents' polarization~\cite{amelkin2017polar}, their conformity~\cite{das2014modeling} or the effect of the influence of opinion leaders within the population~\cite{zhao2018understanding}. In addition, there are models that take into account the structure of the social network in the context of opinion formation~\cite{wu2004socialstructureopinionformation}. Relevant to this research is also consideration of factors such as connection density of a network which affect how opinions converge in the population~\cite{NguyenDynamicsofopinionformation}.

\subsection{Subjective vs. Objective Perceptions in Surveys}
A recurring challenge in opinion modelling lies in the discrepancy between subjective perceptions and objective realities. Brenner and DeLamater~\cite{brenner2016} highlight how self-reported survey data can distort outcomes due to identity-driven biases, such as social desirability and cultural influence. Similarly, Olken~\cite{olken2009} contrasts subjective perceptions of corruption with objective field data, revealing systemic inaccuracies driven by cognitive and contextual factors. These findings underscore the limitations of relying solely on survey-based characteristics for modelling, particularly in capturing the nuanced dynamics of minority groups.

\par Problem of fairness in the context of SNA is complex and multifaceted. It may prove challenging, especially for the problem of opinion dynamics. Therefore, it is important that the dataset selected for the purposes of this experiment contains both the detailed record of agents' demographics and the structure of the social network connecting them. Subsequently, the baseline opinion model should provide state-of-the-art performance when it comes to the accuracy of opinion prediction. The following section describes the models chosen for these purposes and why they match the above requirements.

\section{Dataset and Opinion Model}
This section discusses the datasets used in our experiment. It also contains the results of the analysis of the available records. Please refer to the \ref{appendix:Model_diagram} for a graphical diagram containing information on what data has been utilized. NetSense~\cite{NetSense} is a dataset containing records of phone activity of approximately 200 Notre Dame University students over the span of 3 academic years. Throughout the duration of the experiment, these records were collected together with answers to surveys conducted every semester. Within these surveys, students were asked to answer questions about their background, well-being, and interests. Each of the surveys also contained worldview-related questions (e.g., their stance on euthanasia or marijuana usage). This allows observation of how student opinions fluctuated over this period of time and makes it possible to correlate it with the phone records over the relevant period.
\subsection{Survey Questions}
\label{subsection:survey_questions}
In the CoDiNG model paper \cite{CoDING}, six survey questions from the NetSense dataset were selected for simulation, reflecting various social and political topics. The questions and their shortcodes are as follows:
\begin{itemize}
    \item Euthanasia (\textbf{euthanasia})
    \item Federal spending on social security (\textbf{fssocsec})
    \item Federal spending on welfare (\textbf{fswelfare})
    \item Government's role in job guarantees (\textbf{jobguar})
    \item Marijuana legalization (\textbf{marijuana})
    \item Equal rights progress (\textbf{toomucheqrights})
     \end{itemize}
\ref{appendix:sankey} presents a graphical representation of the distribution of answers to each of these six questions.

This opinion model has run its opinion-prediction simulations on user responses to these same questions, obtaining results better than the Naming Game algorithm. Since predicting when this algorithm will mispredict the correct opinion, the results of the CoDiNG experiment will also be a crucial part of the dataset.

\subsection{Minorities in the Dataset.}
\label{subsection:minoritiesDef}
\par At the center of the concept of Fairness lies the idea of homophily. It is a statistical truth that people are biased towards having relationships with others similar to themselves more frequently than not. That may be attributed to their social bubble containing mostly those of a comparable background, a systemic result of how history shaped their community, or even conscious prejudice that some may hold. The same applies regardless of whether we talk about the majority of the population or different types of smaller subgroups: Two people having a similar set of characteristics increases the likelihood of them belonging to each other's social network. This tendency, in turn, leads to the compartmentalization of these groups and affects how they interact with the rest of society. In the context of SNA, identification of these trends allows observation of the negative effects of this phenomenon and the development of appropriate countermeasures. It is therefore of great importance that we verify if a similar bias is present in the case of the tested opinion model. 

Our goal was to identify subsets of the population for whom the model performs visibly better or worse, regardless of whether they would constitute a traditional minority in the socio-economic sense of the word. While many of the minority groups, as will be defined in this section, are in line with a sociological view of the concept, some groups - e.g., individuals with Facebook non-default privacy settings, are counter to a general understanding of the term. 
The main purpose of such a split was to test the behaviour of the algorithm for different social subgroups and verify if, on top of usual signifiers such as gender or race, there are other attributes that tend to correlate with the model's performance. For a graphical representation of these minorities, please refer to the \ref{appendix:tsne}.
\begin{figure}[htbp]
    \centering
    \includegraphics[width=.8\linewidth]{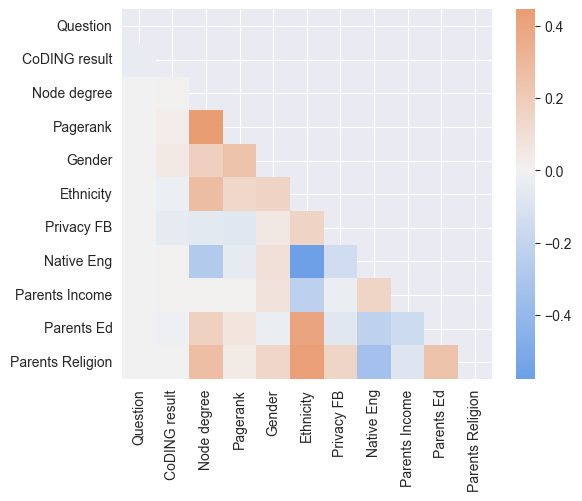}
    \caption{Matrix of correlations between membership in each of the corresponding minorities. Moreover, entries for the node degree, the number of the question, and how well the opinion model algorithm has performed for each of the minorities have been included.}
    \label{fig:minority_correlation_matrix}
\end{figure}
\par Based on the available data, we define groups using the following six features: 

\begin{itemize}
  \item \textbf{Gender} - Within the survey data, 106 people have identified themselves as male and 95 people as female. This splits the population into two groups of, respectively, 52 and 48\% of the population.
  \item \textbf{Ethnicity} - The study enabled students to choose between one of the eight options; White/Caucasian has been selected by more individuals than all the other options combined. Therefore, we have decided to split participants into two groups defined as either White/Caucasian - 137 people - 67\% of the population or Other - 33\%.
  \item \textbf{Facebook privacy} - Though far from any orthodox socio-economic minority, this distinction separates students by how conscious they are of technical aspects of their social media use. An overwhelming majority of the responders have declared their account to be set to the 'All my friends can see my posts' option, whereas only 19\% of participants have either edited this default setting or not provided the answer. This seemingly trivial and not necessarily relevant distinction may contain some valuable insight into how one's level of technical literacy and awareness of computer security may affect their behavioural patterns when it comes to remote communication, and therefore, it has been decided to include this subset within the experiments. 
  \item \textbf{English native} - Amongst the experiment participants 84\% were native to the English language.
  \item \textbf{Parents' income} - Participants were asked to estimate their parents' yearly income level and place it in one of the 14 available brackets. Those varied from less than 10 thousand USD per year to more than 250 thousand USD per year. 27(13\%) students were either not sure of the answer or did not provide one. A threshold of 200 thousand USD has been set as a boundary, splitting the population into a 15\% minority of higher-earning families and 72\% of lower-earning families. This would, according to the data\cite{Census2011ACSDT1Y2011.B19001} from the time of the survey, place such a minority within the fifth percentile of the USA's highest-income households.
  \item \textbf{Parents' education} - Survey provides an insight into the degree of education of students' parents. Participants could declare separately what type of education each of their parents has obtained. It has been chosen to distinguish minorities for individuals for whom neither parent has graduated from college or university. This establishes a minority of around 15\%. As it may be observed in the matrix from the Fig.\ref{fig:minority_correlation_matrix}, membership within this minority correlated most strongly with students' ethnicity and their parents' religious affiliation.
  \item \textbf{Religious affiliation} of one's family has been established as a final of the minority determiners. For that purpose, individuals who have at least one parent not identifying as a Roman Catholic were picked. Within the dataset, a majority of participants - 60\% have both of their parents being members of this religion, whereas the remaining 40\% are from families of a different background (please note that one of the parents of such an individual may still identify as a Roman Catholic).
\end{itemize}

\subsection{Opinion Model}
CoDiNG is a hybrid opinion model based on the Naming Game that incorporates both continuous and discrete components~\cite{CoDING}. The continuous component can be viewed as a representation of the internal state of our mind, reflecting the strength of an individual's opinion. The discrete component can be interpreted as the verbalization of this internal state -- how people simplify complex thoughts into categorical expressions, such as agreeing or disagreeing.
In CoDiNG, each agent $v_i$, similar to the standard Naming Game, holds one of three discrete opinion states -- A, B, or AB -- and is also assigned a two-dimensional opinion vector $o_{v_i} = \langle o_A, o_B \rangle$, where $o_A$ and $o_B$ represent the agent's internal preference strengths for opinions A and B, respectively, both constrained to the interval $[0,1]$. During interactions, agents reveal only their discrete opinions. If the absolute difference between opinion weights, $\Delta_o = |o_A - o_B|$, exceeds a threshold $\gamma$, the expressed opinion corresponds to the higher of the two values -- either A or B. If the difference is less than or equal to $\gamma$, the agent expresses a mixed opinion, denoted as AB. 
We chose to use CoDiNG in our study because it allows for the expression of a third, ambiguous state instead of a strict binary opinion. Moreover, its hybrid nature more closely reflects real-world opinion formation than models relying solely on discrete or continuous approaches. It has also been shown, using real-world data, that CoDiNG often outperforms the classical Naming Game model.

\subsection{Exploratory Data Analysis}
\label{dataanalysis}

\par Our Exploratory Data Analysis was conducted to comprehensively understand the data landscape, characterize the nature of the NetSense study's survey data, and the CoDiNG model's behaviour. Some of the noteworthy insights include:

\subsubsection{Opinion Distribution and Volatility}
For each question, we quantified the proportion of respondents holding the less common stance. We then analyzed `Minority Opinion Rates` for our defined minority subgroups. For example, concerning \textit{euthanasia}, individuals from religious minorities (where at least one parent was not Roman Catholic, "momrelig, dadrelig") held the minority opinion at a rate of 39.3\%, and those who changed Facebook privacy settings ("fbprivacy") did so at 35.5\%. These rates were often substantially higher than for their majority counterparts, indicating a propensity for certain subgroups to hold non-mainstream views. For instance, on \textit{social security}, the "ethnicity" group (non-White/Caucasian) reported a minority opinion rate of 17.2\% compared to 5.1\% for the White/Caucasian majority.
\par
To understand opinion stability, we measured `Volatility (Opinion Changes) by Minority`, defined as the average number of times individuals within each subgroup changed their opinion on a given question across survey waves. Significant differences emerged. For the \textit{job guarantee} question, the "momed, daded" minority (both parents without college-level education) showed an average volatility of 1.57 changes, compared to 1.20 for the majority. For \textit{euthanasia}, this same minority group had a volatility of 1.11 versus 0.67 for their majority counterparts. This highlighted that certain groups exhibited more dynamic opinion trajectories, a potential precursor to CoDiNG misclassification.

\subsubsection{CoDiNG Baseline Misclassification Trends}
\par We assessed `Minority Group Misprediction` rates of the original CoDiNG model. This revealed that for certain questions and minority groups, CoDiNG's baseline error rate was already disproportionately high. For instance, for the \textit{job guarantee} question, the "ethnicity" minority experienced a CoDiNG misprediction rate of 72.9\%, and for \textit{toomucheqrights}, the "parents income" (parental income < \$200k) minority had a misprediction rate of 66.7\%.

\subsubsection{Intersectionality and Misclassification}
\par Recognizing the complexity of social identity, we analyzed `Misprediction by Intersectionality` by CoDiNG. This involved calculating CoDiNG's misprediction rate for individuals based on the number of minority statuses they concurrently held. A clear trend emerged: for \textit{euthanasia}, misprediction rates by CoDiNG rose from 49.2\% for individuals with one minority status to 75.9\% for those with five intersecting minority statuses. Similarly, for \textit{job guarantee}, rates increased from 47.1\% (one status) to 62.1\% (five statuses), indicating a compounding vulnerability.

\subsection{Data and code availability}
In order to ensure the reproducibility of our experiment, transparency and to enable further research in this area, the code, together with the dataset have been made available via the Code Ocean platform via a steady link: 
\begin{center}\boxed{https://doi.org/10.24433/CO.4471647.v1} \end{center} 

\par
This section delved into both the NetSense dataset and the results of the CoDiNG study. We proposed a way of dividing the population into a set of subgroups referred to as minorities. Moreover, we conducted an analysis of the dataset that gives valuable insights into the structure of the population, the interplay between certain attributes, and how well the CoDiNG model has performed, and some tendencies that minorities displayed throughout the experiment. The following section focuses on the applied methodology and technical aspects of the machine learning process used to assess the fairness of the opinion prediction system.

\section{Methodology}
%%% NEW SECTION %%%
% some rationale behind the ML apprach 
% model selection robustness check

%%% REWORK STRUCTURE %%%
% re-iterate the experiment overview
% explain purpose of two-classifier approach
% survey + topology fetures overview
% model training (simple?)
% introduce hybrid 
% segway into the results

The objective of this study is to detect systematic misclassification by the CoDiNG opinion-dynamics model and determine which user attributes—demographic, behavioural, or structural—correlate with these errors. Using the NetSense dataset, which contains semester-wise surveys and detailed communication-network data, we analyze how opinions change for different groups of people over time. We further examine whether certain groups experience disproportionately poor predictive performance and whether such errors can be anticipated from individual-level features.
%The goal of the experiment was to detect any potential discrimination on the side of opinion model and, if it existed, determine with what set of agent's attributes does it coincide. To that end, a set of NetSense surveys has been utilized. Throughout the duration of that experiment, participants answered not only the questions pertaining to their personal wellbeing and social activities, but also a set of worldview-related questions. This collection of repeatedly updated opinions, together with the detailed information about one's background and social interactions conducted in that time, provides a truly unique opportunity to observe how opinions change for different groups of people and to find out if any of these groups suffer from a worse algorithmic performance as compared to others.

\subsection{Experimental Objective}

We focused on six worldview-related questions in specific - monitoring student stances on: euthanasia, social security, welfare, job guarantee, marijuana use, and equal-rights expansion—each answered up to six times over the NetSense study period (see Section 4.1 for question details). These longitudinal opinion traces enable us to evaluate whether CoDiNG assigns correct discrete opinions at each survey wave. For every user–question pair, we construct a binary target variable indicating whether CoDiNG predicted the opinion correctly for each of these six worldview-related questions.%and whether there are presently too many equality-oriented rights. Participants have answered them every semester for the duration of the experiment, resulting in up to six updates for each person. A detailed overview of these questions has been included in Subsection \ref{subsection:survey_questions}.
%While the CoDiNG model offers a better performance than the Naming Game algorithm when it comes to the problem of opinion prediction, it is still not able to correctly estimate all cases, therefore creating an opportunity to examine correlation between attributes of different user-groups and algorithmic performance. 
Therefore, for each of the users, we tried to predict whether the model would assign their opinions correctly for each of these six worldview-related questions. 

We train interpretable machine learning based classifiers that can be used to analyze which attributes are predictive of misclassification and how this varies across minority subgroups. This allows us to compare the predictive contribution of demographic versus topological information.

Once transparent classifier algorithms are trained, it is possible to obtain insight into which training features are relevant while predicting for whom the opinion algorithm will be incorrect, and compare the importance of socio-demographic and topological features.

\subsection{Classifier Objective}%{Objective of the Classifiers}

We train interpretable tree-based classifiers to predict which users are likely to be misclassified by CoDiNG. The key questions are:

\begin{enumerate}
    \item \textit{Is there a learnable pattern in CoDiNG errors, particularly for minority groups?} Specifically, this study investigates whether prediction accuracy for minorities differs significantly from that of the general population, thus identifying potential biases in the performance of the opinion model.
    \item \textit{Do survey-based features, topological features, or their combination offer better predictive power with lower misclassification?} Misclassification refers to discrepancies between ground-truth opinions and simulated opinions generated by the CoDiNG model. 
    %\item Do precision–recall differences (via F1 score) between subgroups indicate potential algorithmic bias?
\end{enumerate}

%Our approach focuses on what insights into the problem of fairness one may obtain from analysis of the interpretable classifiers trained to predict which individuals are going to be assigned opinions incorrectly.
%\par In order to assess this, tree-based classifiers are utilized to recognize whether there is a learnable pattern of misclassification for entities identified as minorities. Specifically, this study investigates whether predictive precision and recall (measured by the F1 score) for minority subgroups differ significantly from that of the general population, thus identifying potential biases in the performance of the model.
%\par A secondary objective is to evaluate whether the incorporation of structural (topological) and attribute-based (survey) characteristics improves the F1 score of  misclassification prediction. Misclassification refers to discrepancies between ground-truth opinions and simulated opinions generated by the CoDiNG model. 

\subsection{Feature Engineering} % for Misclassification Prediction

Because we evaluate survey-only, topology-only, and hybrid models, we do the feature construction for these three pipelines.
As the NetSense experiment progressed over the period of three academic years, some participants quit or stopped responding to the surveys, therefore slightly reducing the available training data; however, profiles with partial survey completion were still included where possible.

\subsubsection{Survey-Based Features}

The survey-based model operates on the data that students have provided in periodic surveys (each semester). Survey features include demographic traits (gender, ethnicity, income, parental education), behavioural attributes, psychological well-being, academic and social engagement metrics, and over 200 self-reported variables collected each semester. These features capture individual preferences, backgrounds, and worldviews, enabling us to test whether the errors in an opinion model (CoDiNG) correlate with specific demographic or socio-economic characteristics.

\subsubsection{Network Topology-Based Features}
Topology features are derived solely from the communication network reconstructed from the phone records provided by the original experiment smartphone logs. These features include node degree, average neighbour's node degree, various centrality measures such as degree, betweenness -\cite{beetweenness_centrality}, closeness -\cite{closeness}, load -\cite{LoadCentrality}, eigenvector -\cite{eigenvector_centrality}, current flow, information -\cite{currentflowAndInformation}, subgraph -\cite{subgraph} and laplacian -\cite{laplacian}; pagerank measure together with an aggregate of student's activity expressed by the CogsNet weight together with the average value of this metric for all of the agent's neighbours. These capture a user’s position in the social graph and allow us to test whether network structure alone predicts CoDiNG failures, and whether structurally marginal users form topological ``minorities." 

\subsubsection{Hybrid Feature Integration}
As a last step, a hybrid model is constructed. It merges all survey and topological features, allowing us to compare it with the two previous models for each of the cases and determine if there are any observable boosts in performance of the classifier once the whole dataset is being utilized for the task.

\subsection{Data Splitting and Validation Strategy}\label{subsubsection:hyperparameter_tuning}

The dataset is split into 80\% training subsets and 20\% testing subsets, stratified by the target variable to ensure a balanced representation of misclassified and correctly classified samples in both subsets. Model selection uses 10-fold stratified cross-validation, with the F1 score as the evaluation metric due to label imbalance.

\subsubsection{Iterative Feature Subset Evaluation}

To avoid overly large feature spaces, we perform an internal iterative feature-ranking and selection loop:
\begin{enumerate}
    \item Compute feature importances (Gini importance for Random Forests, impurity-reduction for Decision Trees).
    \item Rank all features.
    \item Evaluate progressively larger subsets (top 15, 20, 30, full set).
    \item Retrain the model using tuned hyperparameters.
    \item Select the subset with the highest F1 score.
\end{enumerate}

This process is repeated separately for each model and question, ensuring tailored feature sets and minimizing overfitting.

\subsection{Classifier Models}
To classify whether an opinion is likely to be misclassified by the CoDiNG model, multiple interpretable classifiers were tested. However, for each of the three approaches, the Stratified Random Forest and Decision Tree classifiers consistently obtained the best results.

The results in Table \ref{tab:all_models_results} show that the Stratified Random Forest model performed better in the case of the demography-based and hybrid model, while topology-based approach obtained best results when using a Decision Tree classifier.

\begin{table}[!htbp]
\caption{F1 scores for all three approaches for each question. The best results are in bold.}
\label{tab:all_models_results}
\resizebox{\textwidth}{!}{%
\begin{tabular}{|lllllll|}
\hline
\rowcolor[HTML]{EFEFEF} 
\multicolumn{1}{|l|}{\cellcolor[HTML]{EFEFEF}} &
  \multicolumn{1}{l|}{\cellcolor[HTML]{EFEFEF}Euthanasia} &
  \multicolumn{1}{l|}{\cellcolor[HTML]{EFEFEF}\begin{tabular}[c]{@{}l@{}}Social \\ Security\end{tabular}} &
  \multicolumn{1}{l|}{\cellcolor[HTML]{EFEFEF}Welfare} &
  \multicolumn{1}{l|}{\cellcolor[HTML]{EFEFEF}\begin{tabular}[c]{@{}l@{}}Job \\ Guarantee\end{tabular}} &
  \multicolumn{1}{l|}{\cellcolor[HTML]{EFEFEF}Marijuana} &
  \begin{tabular}[c]{@{}l@{}}Too Much \\ Eq. Rights\end{tabular} \\ \hline
\rowcolor[HTML]{DAE8FC} 
\multicolumn{7}{|c|}{\cellcolor[HTML]{DAE8FC}Demography-based approach} \\ \hline
\multicolumn{1}{|l|}{\cellcolor[HTML]{DAE8FC}Stratified RF} &
  \multicolumn{1}{l|}{\textbf{0.5602}} &
  \multicolumn{1}{l|}{\textbf{0.5708}} &
  \multicolumn{1}{l|}{\textbf{0.5892}} &
  \multicolumn{1}{l|}{\textbf{0.5318}} &
  \multicolumn{1}{l|}{\textbf{0.5757}} &
  \textbf{0.5678} \\ \hline
\multicolumn{1}{|l|}{\cellcolor[HTML]{DAE8FC}Decision Tree} &
  \multicolumn{1}{l|}{0.5150} &
  \multicolumn{1}{l|}{0.5200} &
  \multicolumn{1}{l|}{0.5400} &
  \multicolumn{1}{l|}{0.4850} &
  \multicolumn{1}{l|}{0.5300} &
  0.5250 \\ \hline
\rowcolor[HTML]{FFCC67} 
\multicolumn{7}{|c|}{\cellcolor[HTML]{FFCC67}Topology-based approach} \\ \hline
\multicolumn{1}{|l|}{\cellcolor[HTML]{FFCC67}Stratified RF} &
  \multicolumn{1}{l|}{0.5290} &
  \multicolumn{1}{l|}{0.4380} &
  \multicolumn{1}{l|}{0.4200} &
  \multicolumn{1}{l|}{0.5080} &
  \multicolumn{1}{l|}{0.5090} &
  0.5950 \\ \hline
\multicolumn{1}{|l|}{\cellcolor[HTML]{FFCC67}Decision Tree} &
  \multicolumn{1}{l|}{\textbf{0.5524}} &
  \multicolumn{1}{l|}{\textbf{0.4555}} &
  \multicolumn{1}{l|}{\textbf{0.4410}} &
  \multicolumn{1}{l|}{\textbf{0.5244}} &
  \multicolumn{1}{l|}{\textbf{0.5241}} &
  \textbf{0.6225} \\ \hline
\rowcolor[HTML]{9AFF99} 
\multicolumn{7}{|c|}{\cellcolor[HTML]{9AFF99}Hybrid approach} \\ \hline
\multicolumn{1}{|l|}{\cellcolor[HTML]{9AFF99}Stratified RF} &
  \multicolumn{1}{l|}{\textbf{0.5147}} &
  \multicolumn{1}{l|}{\textbf{0.4839}} &
  \multicolumn{1}{l|}{\textbf{0.6021}} &
  \multicolumn{1}{l|}{\textbf{0.5161}} &
  \multicolumn{1}{l|}{\textbf{0.7327}} &
  \textbf{0.5835} \\ \hline
\multicolumn{1}{|l|}{\cellcolor[HTML]{9AFF99}Decision Tree} &
  \multicolumn{1}{l|}{0.4925} &
  \multicolumn{1}{l|}{0.4600} &
  \multicolumn{1}{l|}{0.5850} &
  \multicolumn{1}{l|}{0.5000} &
  \multicolumn{1}{l|}{0.7100} &
  0.5600 \\ \hline
\end{tabular}%
}
\end{table}

After selecting the best models for each type of feature, the next section will focus on their analysis and identify systematic performance gaps across socio-economic, demographic, behavioural, and topological minorities. We analyse which features drive prediction misclassifications and when and why either demographic or structural information becomes dominant.

\section{Results and Fairness Evaluation}
\label{experiment_results}
\noindent
\begin{figure}
    \includegraphics[width=\linewidth,left=2\textwidth]{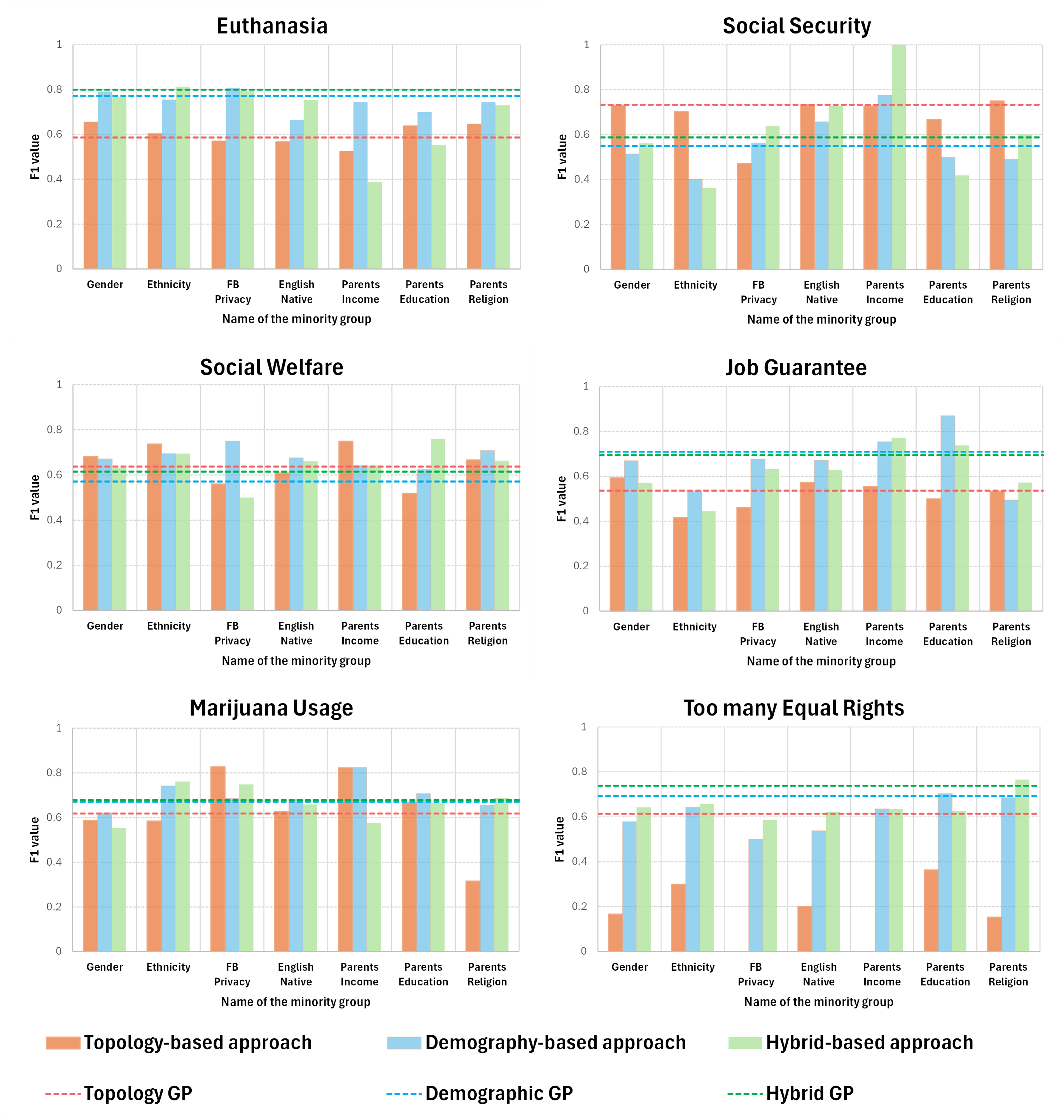}
    \caption{Diagrams of F1 scores for classifiers predicting CoDiNG model misclassifications. Results are shown for each of the six survey questions and across seven identified minority groups. Dashed lines indicate the F1 score for the general population for Survey-Based (Blue), Topology-Based (Red), and Hybrid (Green) approaches.}
    \label{fig:results_column}
\end{figure}

The empirical investigation into predicting CoDiNG model misclassifications reveals substantial performance differences across the Survey-Based, Topology-Based, and Hybrid modelling approaches, as illustrated in Figure~\ref{fig:results_column}. A salient overarching observation is the significant influence of survey question characteristics on model efficacy. Drawing from the original NetSense study~\cite{NetSense}, questions concerning \textit{euthanasia}, \textit{marijuana} usage, and \textit{job guarantees} typically elicited polarized or consensus-driven responses. Conversely, topics such as \textit{social security}, \textit{welfare}, and attitudes towards \textit{equal rights} were marked by a greater degree of uncertainty or indifference among respondents. This distinction is paramount, as our findings indicate that opinions on topics fraught with uncertainty (e.g., welfare, social security) were often more accurately predicted using topological features. This underscores the amplified role of social structure in contexts where individual conviction may be less potent~\cite{holme2015}, a theme that recurs when examining performance for specific minority subgroups.

Following hyperparameter optimization and feature selection, the Stratified Random Forest classifier was identified as optimal for the Survey-Based and Hybrid approaches, while the Decision Tree Classifier was selected for the Topology-Based approach, based on their F1 scores for the general population (Table \ref{tab:all_models_results}). The subsequent analysis meticulously examines the performance of these chosen classifiers, with a particular focus on their capacity to predict misclassifications for distinct minority groups across the diverse spectrum of opinion questions.

\par The subsequent analysis meticulously examines the performance of these chosen classifiers. It is important to recall from our EDA (Section~\ref{dataanalysis}) that the CoDiNG model itself exhibited varying baseline misclassification rates across subgroups, often linked to factors like opinion volatility and intersectionality. Our fairness classifiers aim to predict these pre-existing patterns of misclassification.

\subsection{Minority Subgroup Performance}
The inherent nature of the survey question (i.e. whether it addresses a topic of Consensus, Polarization, or Apathy) profoundly influences which modelling strategy is most effective for predicting misclassifications, especially for minority subgroups.
\begin{itemize}
    \item \textbf{Consensus Topics} (e.g. \textit{Euthanasia, Job Guarantee}): Survey-based models generally excelled for the overall population, with Hybrid models offering modest improvements. However, even with hybrid features, some minority groups (e.g., Parents' income and parents' education on \textit{Euthanasia}; Ethnicity on \textit{Job Guarantee} with the Hybrid model) remained less predictable.
    \item \textbf{Polarized Topics} (e.g. \textit{Marijuana}): Survey-based models again showed strong performance. Hybrid model efficacy for the general population was comparable or sometimes slightly inferior. Crucially, for certain minorities (e.g., Parents' Income), the Survey-based approach provided notably superior predictions.
    \item \textbf{Apathetic Topics} (e.g. \textit{Social Security, Welfare, Equal Rights}): These topics presented the most complex interaction effects. Survey-based models exhibited lower F1 scores for the general population, yet specific minorities (e.g., Parents' Income on \textit{Social Security}) achieved high predictability using only survey attributes. These questions also distinctly showcased the niche strengths of Topology-based models for other demographic minorities. The Hybrid model often improved overall scores and achieved remarkable results for some (e.g., Parents' Income on \textit{Social Security}), but its dominance was not uniform, with Survey or Topology approaches sometimes proving more effective for particular subgroups.
\end{itemize}
This overarching pattern suggests that the underlying cognitive and social processes of opinion formation and expression vary across issue types, thereby altering which feature sets (individual attributes, network structure, or their combination) are most indicative of potential model misclassifications for vulnerable populations~\cite{holme2015, dellaposta2015}.

Minority groups defined by specific behaviours, such as the "FB Privacy" cohort (those altering default online privacy settings), demonstrated extreme fluctuations in the predictability of their misclassification, contingent on the modelling approach.
\begin{itemize}
    \item Utilizing the Survey-based model, this group achieved high F1 scores on several questions (e.g., 0.806 for \textit{Euthanasia}, 0.750 for \textit{Welfare}), indicating that their survey-reported attributes are potent predictors of misclassification in these instances.
    \item In stark contrast, the Topology-based model performed exceedingly poorly for this group on other questions (e.g., F1 $\approx$ 0.000 for \textit{toomucheqrights}, as suggested by Figure~\ref{fig:results_column}). This aligns with this study's broader observation of topological model failure for structurally isolated entities~\cite{jackson2019}.
    \item The Hybrid model offered no consistent resolution to these extremes, performing well for this group on \textit{Euthanasia} (F1=0.806) and \textit{Marijuana} (F1=0.746), but substantially underperforming the Survey-based model on \textit{Welfare} (Hybrid F1=0.500 vs. Survey F1=0.750) and struggling on \textit{toomucheqrights} (Hybrid F1=0.590).
\end{itemize}
This high variance implies that for minorities whose defining characteristics are behavioral and may correlate with network phenomena like structural isolation, their individual survey responses might capture strong, predictive personal stances, while their network data could be either too sparse, too unique, or insufficiently reflective of typical social influence dynamics to be useful.

Minorities distinguished by parental socio-economic indicators ("Parents' income", "Parents' education") or familial religious affiliation ("Parents' religion") consistently demonstrated that these background characteristics are strong predictors of CoDiNG model misclassifications. However, the optimal modelling strategy to leverage these potent signals varied considerably across different opinion questions.
\begin{itemize}
    \item The "Parents' income" minority, for example, attained high F1 scores with Survey-based models on \textit{Marijuana} (0.825) and \textit{Social Security} (0.775), and a perfect F1 score of 1.000 with the Hybrid model on \textit{Social Security}. Yet, for this same group, Hybrid model performance was notably inferior for \textit{Euthanasia} (F1=0.378) and \textit{Marijuana} (F1=0.571), where the Survey-only approach was more effective.
    \item The "Parents' education" minority showed exceptional F1 scores (0.869) with the Survey-only model on the \textit{Job Guarantee} question, a performance level that diminished when using the Hybrid model (F1=0.733).
\end{itemize}
This pattern suggests that deeply ingrained familial and socio-economic background factors, which are often robustly captured in survey data, can exert direct, powerful, yet context-dependent influences on opinion formation pathways that subsequently lead to model misclassifications. The variable success of the Hybrid model implies that the interplay between these strong background factors and network topology is complex and not always synergistically additive for predictive purposes.

\subsubsection{Survey-Based Performance Insights}
The Survey-Based approach, leveraging self-reported demographic and attribute data, demonstrated considerable strength in predicting CoDiNG misclassifications for the general population, especially when opinions were clearly defined or reflected strong personal conviction. This was evident in its robust F1 scores for consensus topics such as \textit{Euthanasia} (F1 = 0.771) and \textit{Job Guarantee} (F1 = 0.709). An analysis of feature importance for the \textit{Job Guarantee} question revealed "ethnicity" and survey-derived social network metrics (e.g., Community, Betweenness Centrality based on self-reported friendships) as highly influential predictors.
Beyond the general population, the Survey-Based approach exhibited notable efficacy for specific minority subgroups, occasionally achieving exceptional F1 scores that surpassed its general performance levels:
\begin{itemize}
    \item For the \textit{Job Guarantee} question, this approach yielded a striking F1 score of 0.869 for the "Parents' education" minority.
    \item On the \textit{Euthanasia} question, the "FB Privacy" minority demonstrated a high F1 score of 0.806.
    \item Even on "apathetic" topics where its general performance was lower, the Survey-Based model showed strength for certain minorities. For \textit{Social Security} (overall F1 = 0.549), the "Parents' income" minority reached an F1 of 0.775. Similarly, for \textit{Welfare} (overall F1 = 0.571, per detailed results), the 'FB Privacy' minority achieved an F1 of 0.750.
    \item For the polarized topic of \textit{Marijuana} (overall F1 = 0.671, per detailed results), the "Parents' income" minority showed high predictability with an F1 of 0.825.
\end{itemize}
These instances underscore that for particular minority groups, their individual attributes and self-perceptions, as captured by survey data, can serve as remarkably strong and direct indicators for predicting opinion model misclassifications. However, the model's overall accuracy for the general population tended to decrease for topics characterized by widespread uncertainty or low engagement, such as \textit{Social Security} and \textit{Welfare}, aligning with the challenges identified by DellaPosta et al. (2015)~\cite{dellaposta2015} in modelling opinions under such conditions.
\subsubsection{Topology-Based Performance Insights}
This model's principal utility becomes apparent in specific analytical contexts, most notably in predicting misclassifications for "apathetic" opinion questions where social influence dynamics might supersede individual conviction as primary drivers.

Crucially, this approach demonstrated unique and sometimes superior predictive power for certain demographic minorities when addressing these apathetic topics, occasionally outperforming both Survey-Based and even Hybrid models for these specific subgroup-question combinations:
\begin{itemize}
    \item For the \textit{Social Security} question, the 'Gender' minority achieved a noteworthy F1 score of 0.730. Visual inspection of Figure~\ref{fig:results_column} further suggests that the Topology-Based model performed strongly for "Ethnicity" and "Parents' religion" minorities on this topic, relative to their performance with the Survey-Based model.
    \item For the \textit{Welfare} question, the "Ethnicity" minority attained an F1 score of 0.737 with the topology model.
\end{itemize}
These results suggest that when individual opinions are less crystallized or personally salient, an individual's position within the social fabric and their patterns of interaction become more critical predictors of how their views might diverge from model expectations and thus be misclassified.

Despite these niche strengths, the Topology-Based model exhibited severe limitations for subgroups characterized by structural isolation. A stark example is the 'FB Privacy' minority, for which this model yielded an F1 score approaching 0.000 when predicting misclassifications on the \textit{toomucheqrights} question. This finding is consistent with Jackson's (2019)~\cite{jackson2019} research on structural isolation posing a significant barrier to accurate network-based modelling and underscores that purely topological methods are ill-suited for understanding subgroups that are detached from mainstream network interactions or influence flows. Confirmatory statistical analyses (Kruskal-Wallis H-test, as mentioned in the original draft) further indicated the detrimental impact of network isolation on predictive efficacy for this model.
\subsubsection{Hybrid Model Comparative Analysis}
The Hybrid model, which synergistically integrated survey-derived demographic attributes with topology-based network metrics, generally provided the most consistent and often the highest F1 scores for predicting misclassifications for the general population across the spectrum of opinion topics. It achieved overall F1 scores of 0.795 for \textit{Euthanasia}, 0.589 for \textit{Social Security}, 0.605 for \textit{Welfare}, and 0.727 for \textit{toomucheqrights} (based on detailed results from the appendix). This general outperformance aligns with calls for integrative modelling approaches, such as that by Barberá (2020)~\cite{barbera2020}.

This approach frequently led to improved predictive accuracy for many minority subgroups when compared to single-modality models, with some instances showing dramatic gains:
\begin{itemize}
    \item On the \textit{Social Security} question, the "Parents' income" minority achieved a perfect F1 score of 1.000, demonstrating exceptional predictability with the combined feature set.
    \item For \textit{Euthanasia}, the "Ethnicity" minority reached an F1 score of 0.810.
    \item For the \textit{Toomucheqrights} question, the "Parents' religion" minority demonstrated an F1 score of 0.742.
\end{itemize}
As illustrated by the dashed lines in Figure~\ref{fig:results_column}, while the performance uplift from the Hybrid model for the general population was sometimes marginal compared to the best single-modality approach, the positive impact was often notably more pronounced and beneficial for specific minority segments.

However, the Hybrid model's enhanced performance was not universal across all minority groups and opinion questions. Several critical instances emerged where the Survey-only model provided superior misclassification predictions for particular minorities:
\begin{itemize}
    \item For the "Parents' income" minority, Survey-based F1 scores were substantially higher for \textit{Euthanasia} (Survey: 0.743 vs. Hybrid: 0.378) and \textit{Marijuana} (Survey: 0.825 vs. Hybrid: 0.571).
    \item The "Parents' education" minority showed higher predictability with the Survey-based approach for \textit{Job Guarantee} (Survey: 0.869 vs. Hybrid: 0.733), as well as for \textit{Euthanasia} and \textit{toomucheqrights}.
    \item For the "FB Privacy" minority, the Survey-based F1 score was markedly higher for the \textit{Welfare} question (Survey: 0.750 vs. Hybrid: 0.500).
\end{itemize}
These exceptions highlight an important nuance: while feature integration is generally a powerful strategy, it can occasionally dilute an exceptionally strong and clear predictive signal present in attribute-based data, particularly for certain subgroups whose defining characteristics might be overwhelmingly captured by survey responses for specific topics.

\section{Discussion}
\par The findings of this study, focused on predicting misclassifications by the CoDiNG opinion dynamics model, offer a multifaceted understanding of algorithmic fairness in the context of event-based social data. Our analysis, leveraging survey-based, topology-based, and hybrid modelling approaches, reveals that the path to equitable opinion prediction for minority subgroups is not monolithic.

Instead, it is profoundly shaped by an interplay of individual attributes (often captured by surveys), network structures (derived from interactions), the inherent characteristics of the opinion topic, and the specific nature of the minority group itself. Our detailed EDA further uncovered that factors like opinion volatility, the propensity to hold minority opinions, and intersectional identity significantly influence the CoDiNG model's baseline performance, creating systemic patterns of error that our fairness classifiers subsequently attempt to predict. This section interprets these complex findings, connects them to broader theoretical considerations, and outlines their implications.
\subsection{Understanding CoDiNG's Misclassification Patterns}
\par Against this backdrop of CoDiNG's inherent differential performance, we developed three types of fairness classifiers (Survey-Based, Topology-Based, and Hybrid) to predict these misclassifications.
There are notable differences in their performance depending on the nature of the survey question. As a general trend, the Survey-Based approach often performed strongly for the general population, particularly on consensus or polarized topics. However, the Topology-Based model demonstrated unique strengths, particularly for certain demographic minorities on "apathetic" topics like \textit{Welfare} and \textit{Social Security}. This aligns with the notion that when individual conviction is low, and opinions might be more volatile or influenced by immediate social context, network structure becomes a more salient predictor of CoDiNG's errors~\cite{holme2015}.
\subsubsection{Key Patterns in Model Performance}
\label{subsubsection:patterns}
As a general trend, the Survey-Based approach often performed strongly for the general population, particularly on consensus or polarized topics. However, the Topology-Based model demonstrated unique strengths, particularly for certain demographic minorities on "apathetic" topics like \textit{Welfare} and \textit{Social Security}. This aligns with the notion that when individual conviction is low, and opinions might be more volatile or influenced by immediate social context, network structure becomes a more salient predictor of CoDiNG's errors~\cite{holme2015}.

Specifically, concerning minorities and our fairness classifiers:
\begin{itemize}
    \item Survey-Based models showed particular strength for certain minorities on topics like \textit{Euthanasia} (e.g., 'FB Privacy' F1=0.806; "Parents' income" F1=0.743) and \textit{toomucheqrights} (e.g., "Parents' education" F1=0.704). On \textit{Job Guarantee}, the "Parents' education" minority reached an exceptional F1 of 0.869 with this approach. This suggests that for these minority-question pairs, individual attributes or background factors captured by surveys were paramount in predicting CoDiNG's misclassifications.
    \item Topology-Based models excelled for other specific minority-question combinations, most notably for demographic minorities on apathetic topics. For instance, on \textit{Social Security}, the 'Gender' minority (F1=0.730) and, as suggested by Figure~\ref{fig:results_column}, "Ethnicity" and "Parents' religion" minorities, had their CoDiNG misclassifications better predicted by topology. On \textit{Welfare}, the ethnic minority (F1=0.737) also benefited from the topology-based classifier.
    \item Performance was often similar (difference in F1 $<$ 0.15 between the best single-modality and the hybrid, or between two single-modality approaches) for questions like \textit{Welfare}, \textit{Job Guarantee}, and \textit{Marijuana} when considering the best overall approach for minorities, but with significant variation for individual minority groups within these questions.
\end{itemize}
\par It is indeed worth noting the correlation between the question being perceived as polarizing by respondents and the Survey-Based classifier performing well in predicting misclassifications. Topics for which the demographic approach performed better may be more deeply "grounded in one's world view," with opinions being less susceptible to immediate peer interactions and more reliant on stable personal background characteristics, characteristics that also seem to relate to lower opinion volatility for some groups. In contrast, questions for which the Topology-Based algorithm showed higher F1 scores (particularly for certain minorities on apathetic questions where opinion volatility might be higher or more network-dependent) might indeed be "less politically divisive because of their more abstract nature," or, as with \textit{Marijuana} opinions could be more actively discussed and negotiated within peer groups, making interaction patterns relevant, though not always decisively superior to attribute data for predicting CoDiNG's errors. The cannabis decriminalization debates coinciding with the experiment offer a plausible explanation for heightened network influence or discussion on that particular topic, making topological features more pertinent for predicting misclassifications.
\subsubsection{Influence of Question Typology}
\par  
Any work dealing with topics such as fairness, minority, or subjective opinions is prone to a number of potential biases stemming from a multitude of factors such as the selection of the survey group, subjective bias of the experiment participants in regard to how they interact with the surveys' questions and, of course, the very phrasing of consecutive queries and the amount of workload they presented to the participants. It is therefore not outside of the realm of possibility that a study similar to the NetSense, conducted at a similar time within a relatively similarly constituted group of students, would bear quite different results in regard to how they have answered the presented questions.

\subsubsection{Predictive Signals in Socioeconomic/Religious Minorities}
Minorities distinguished by parental socio-economic indicators ("Parents' income", "Parents' education") or familial religious affiliation ("Parents' religion") consistently demonstrated that these background characteristics are strong predictors of CoDiNG model misclassifications. However, the optimal modelling strategy to leverage these potent signals varied considerably across different opinion questions.
\begin{itemize}
    \item The "Parents' income" minority, for example, attained high F1 scores with Survey-based models on \textit{Marijuana} (0.825) and \textit{Social Security} (0.775), and a perfect F1 score of 1.000 with the Hybrid model on \textit{Social Security}. Yet, for this same group, Hybrid model performance was notably inferior for \textit{Euthanasia} (F1=0.378) and \textit{Marijuana} (F1=0.571), where the Survey-only approach was more effective.
    \item The "Parents' education" minority showed exceptional F1 scores (0.869) with the Survey-only model on the \textit{Job Guarantee} question, a performance level that diminished when using the Hybrid model (F1=0.733).
\end{itemize}
This pattern suggests that deeply ingrained familial and socio-economic background factors, which are often robustly captured in survey data, can exert direct, powerful, yet context-dependent influences on opinion formation pathways that subsequently lead to model misclassifications. The variable success of the Hybrid model implies that the interplay between these strong background factors and network topology is complex and not always synergistically additive for predictive purposes.
\subsubsection{Persistent Challenges for Certain Demographics}
Despite the general advancements offered by the Hybrid approach, some demographic minorities, notably the "Ethnicity" subgroup, consistently presented greater challenges in accurately predicting CoDiNG model misclassifications.
 \begin{itemize}
  \item For instance, the ethnic minority registered Hybrid F1 scores of 0.364 on \textit{Social Security} and 0.433 on \textit{Job Guarantee}. These scores were typically below the general population's performance level and often lower than those of other minority subgroups for the same questions.
 \end{itemize}
 This persistent difficulty, even when employing an integrated feature set, may be linked to underlying factors such as the consistently higher `Opinion Volatility` observed in the "Ethnicity" group (e.g., 1.27 changes on average for \textit{job guarantee} vs. 1.26 for the majority, but also notably high raw misprediction by CoDiNG at 72.9\% for this group on this question as per our EDA in Section~\ref{dataanalysis}), or more intricate forms of `Structural Isolation` not fully mitigated by the combined features. These characteristics are mentioned in the broader literature on fairness in social networks~\cite{karimi2022minorities, karimi2018homophily} and align with our initial findings of differential rates of change of opinion and misprediction of CoDiNG among demographic groups.
\subsubsection{Minority Opinion Holding}
Demographic minorities consistently adopted the less popular stances ("minority opinions") at higher rates across all issues. For instance, on \textit{euthanasia}, individuals from religious minorities (momrelig, dadrelig) held the minority opinion at a rate of 39.3\%, and those who altered their Facebook privacy settings (fbprivacy) did so at 35.5\%. This contrasts with overall lower rates for majority groups (e.g., the majority counterpart for momrelig, dadrelig on euthanasia held minority opinions at 20.7\%). This tendency for certain minorities to hold views divergent from the general consensus presents an intrinsic challenge for opinion models like CoDiNG.
\subsubsection{Opinion Volatility}
Demographic minorities often exhibited substantially higher rates of opinion change across survey waves. For example, on the \textit{job guarantee} question, the momed, daded minority (parents with neither having a college education) showed an average opinion volatility of 1.57 changes per individual, compared to 1.20 for their majority counterparts. Similarly, for \textit{euthanasia}, this same minority group had a volatility of 1.11 versus 0.67. This frequent opinion shifting was associated with higher baseline misclassification rates by the CoDiNG model, as it often failed to anticipate these dynamic transitions, particularly for these subgroups.
\subsubsection{Network Isolation Effects}
As noted in prior literature and supported by our network visualizations (Fig.~\ref{fig:minorities_networks}), demographic minorities sometimes exhibit different network centrality patterns, potentially indicating more peripheral positions. For instance, the "Parents' income" minority (parental income < \$200k) showed significantly lower eigenvector centrality (approx. 0.003 vs. approx. 0.02 for the non-minority on average across individuals in these groups). This structural positioning can correlate with increased misclassification by CoDiNG. For example, this "Parents' income" group had a baseline CoDiNG misclassification rate of 66.7\% on the \textit{toomucheqrights} issue, compared to an overall misclassification rate of around 55\% for the general population on that question.
\subsubsection{Intersectional Vulnerability}
Intersectional demographic minorities (individuals simultaneously belonging to multiple minority groups) experience significantly elevated predictive misclassification rates from the original CoDiNG model. For instance, concerning \textit{euthanasia}, CoDiNG's misclassification rate for individuals with a single minority status was 49.2\%, but this soared to 75.9\% for those with five intersecting minority statuses. This demonstrates that the challenge of accurate opinion prediction by CoDiNG is markedly amplified for individuals at multiple social margins, meaning our fairness classifiers are tasked with predicting errors that are already more frequent and pronounced for these individuals. Future fairness analyses should explicitly model and test classifier performance for intersectional groups.
\subsection{Implications for Fairness Classifier Design}
The comprehensive analysis of experimental results points to several pivotal trends that carry significant implications for model selection in the pursuit of fair opinion prediction.

First, while the Hybrid modelling strategy generally provides the most robust and often the highest predictive accuracy for the general population—and frequently enhances predictability for many minority subgroups—its superiority is not universal. The critical takeaway is the paramount importance of context, defined by the specific attributes of the minority group under consideration and the intrinsic nature of the opinion topic (consensus, polarized, or apathetic).

Second, single-modality approaches retain indispensable value in specific scenarios. The Survey-Based model can be exceptionally powerful and may even outperform hybrid alternatives when strong demographic markers or self-reported attributes are the dominant drivers of opinion formation (or its misclassification) for a particular subgroup on a given topic. Conversely, the Topology-Based model, despite its moderate overall efficacy and pronounced limitations with structurally isolated groups, demonstrates unique and vital strengths. It excels in predicting misclassifications for certain demographic minorities, particularly concerning topics characterized by widespread uncertainty or indifference, where the architecture of social networks may exert a more decisive influence on opinion dynamics than individual predispositions.

Third, the diverse performance profiles across different minority groups and question types underscore the intricate challenge of achieving uniformly equitable algorithmic predictions. No single modelling approach emerged as a panacea guaranteeing the fairest outcomes for all subgroups in all situations. This highlights an urgent need for nuanced, context-aware strategies in the domain of fairness-oriented social network analysis. Factors such as structural isolation and potentially higher opinion volatility within certain minority populations continue to present formidable challenges, even for sophisticated hybrid models, suggesting that these characteristics may require more specialized modelling considerations or feature engineering in future research.

While in the field of Machine Learning, the problem of Fairness has been relatively thoroughly established, it is still an emerging field when it comes to Social Network Analysis. The main difference is the wealth of data provided to the models by including information derived from the structure of network graphs. Although the scope of this work may be narrowed down to just one dataset and a classifying model, its results already indicate quite a deep nuance when it comes to the interplay between one's demographic and topological attributes and how these correlate with the algorithm's performance. Among key findings is the fact that neither the socio-economic features nor the network structure-derived ones are enough on their own to identify agents in danger of being discriminated against by the model. While the exact effect appears to be predicated on the specific type of minority, there is, undoubtedly, a strong discrepancy between how easy it is to detect mispredictions within the general population and within subgroups, which does not seem to correlate with the size of the said subgroup. How effective a given type of classifier is also depends heavily on the kind of question asked. A division into 3 types of questions proposed in this work provides a useful insight into the context of the NetSense dataset; however, further analysis of other available data will be required here. 

\subsection{Recommendations for Future Modelling}
The observed patterns collectively argue against a one-size-fits-all approach to model selection when fairness for minorities is a primary concern. Instead, a more adaptive framework, potentially involving context-specific model choices or ensemble methods, may be necessary to navigate the complex landscape of opinion dynamics and algorithmic bias. Our work focused solely on one dataset and one opinion model; extending to additional models may further improve the level of comprehension and reliability when it comes to a fairness-oriented framework for opinion dynamics.
\par This concludes the discussion section in which we presented the results of our experiment together with the insights gained in the context of performance of the opinion prediction for different minorities and questions. The following sections consist of our conclusions, where we summarize the findings of this paper, look over the contributions it offers to the field of Fairness in the context of SNA, and consider possible future avenues for continued research.

\section{Conclusion}
The problem of Fairness in Social Network Analysis is still an emerging field where much remains to be addressed. This work proposes a unique way of approaching the problem of evaluating fairness in opinion prediction. We looked into how socioeconomic and network structure-derived factors correlate with one's chances of being more often misclassified by the opinion model. Our results suggest a non-trivial interplay of various aspects, such as one's membership in more than one group of a minority status or correlation between the nature of the asked question and the family background of the respondent. It also indicates that while, on average, it is the demographic set of attributes that holds more predictive power in that regard, there are cases (like apathetic questions) where it is the topological features that may prove more informative of that. Moreover, it shows that while, in general, the hybrid model does score the highest, there exist some edge cases where inclusion of both types of data may actually move the focus away from the most optimal solution.

\section{Acknowledgements}
This work was supported by the National Science Centre, Poland, under Grant 2021/41/B/HS6/02798, by the Polish Ministry of Education and Science within the program "International Projects Co-Funded", and by the European Union through Horizon Europe (OMINO) under Grant 101086321

\newpage
\appendix
\section{Diagram of the model's workflow}
\label{appendix:Model_diagram}
\par The following diagram's purpose is to visualize the process of our experiment and to give a better intuition as to which part of the dataset has been used for each of the classifier models.
\begin{figure}[htbp]
    \centering
    \includegraphics[width=\linewidth]{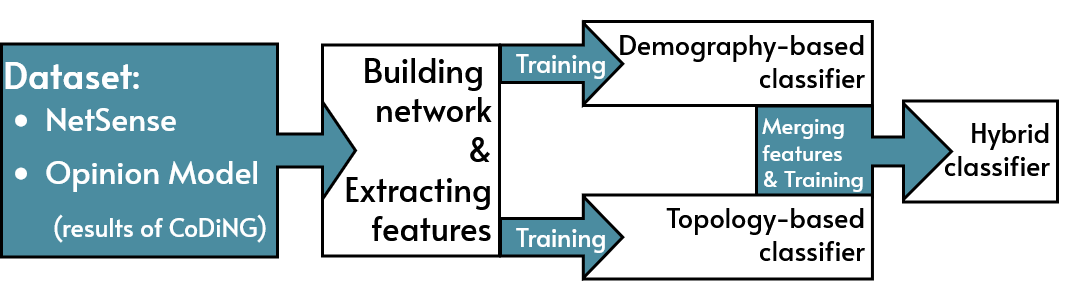}
    \caption{Diagram of the workflow together with the structure of the data used for training of each of the classifier models.}
    \label{fig:WorkFlowDiagram}
\end{figure}

\section{Graphical representation of the minorities within the NetSense dataset}
\par
\begin{figure}[htbp]
    \centering
    \includegraphics[width=0.85\linewidth]{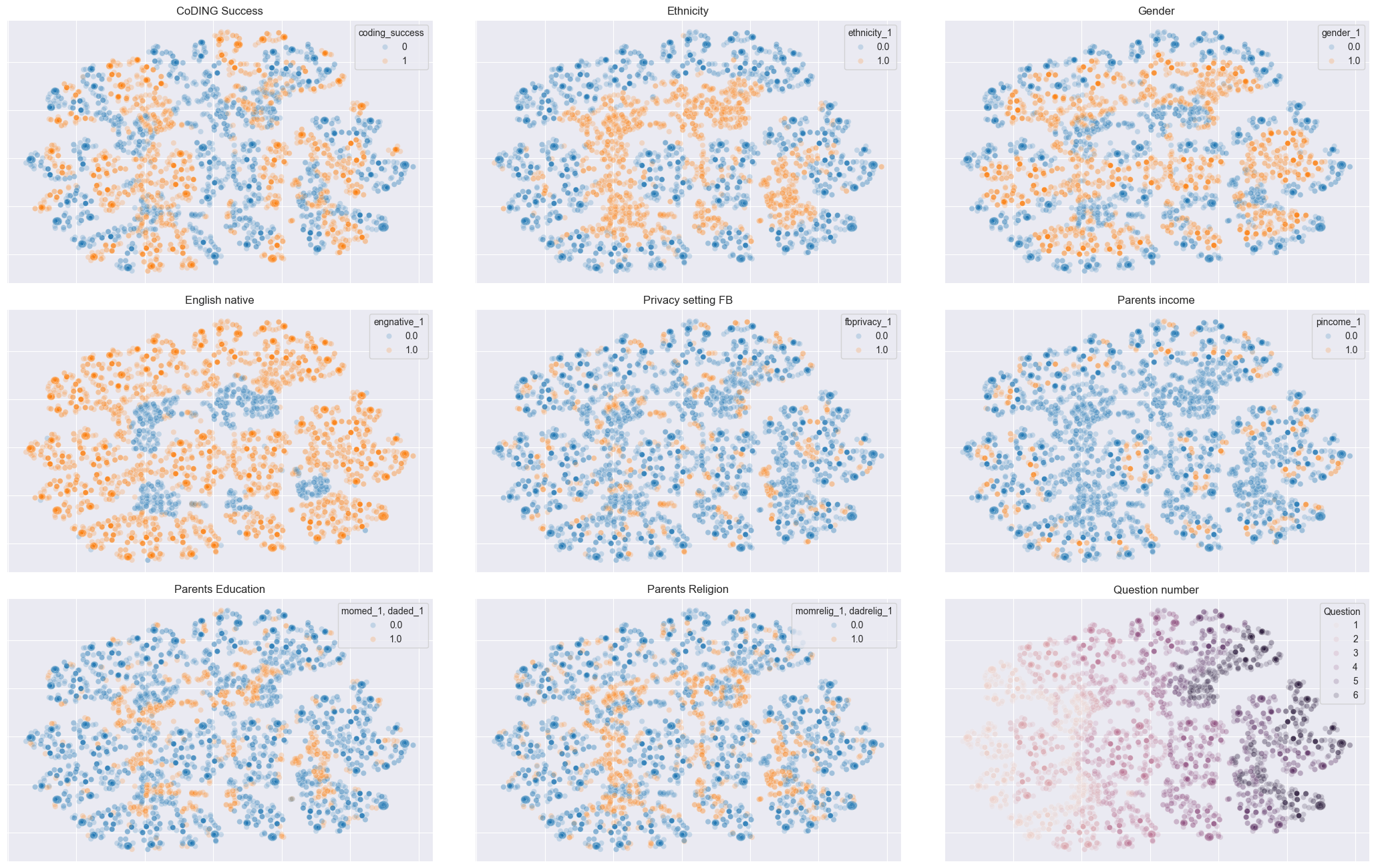}
    \caption{Graphs obtained via the t-SNE dimension reduction method illustrating the distribution of the agents in relation to a set of their chosen characteristics.}
    \label{fig:tsne_demographic_graphs}
\end{figure}

 A visualization of the dataset in Fig.~\ref{fig:tsne_demographic_graphs} illustrates some of the chosen characteristics of the profile of the population. It has been created using the t-SNE method, placing each of the agents in proximity to similar individuals. In the first image (top-left), the results of the CoDING labelling can be seen - orange colour has been used to mark these data points for which the CoDING model performed a successful labelling, otherwise they have been marked blue. 
\par Some overlaps between different minorities may be easily identified here (e.g., quite a strong overlap between Ethnicity and English Native groups). The last of the nine subgraphs is colour-coded in regard to the six worldview questions for which the CoDiNG model tried to predict respondents' answers. 
\label{appendix:tsne}

\begin{figure}
    \centering
    \includegraphics[width=0.85\linewidth]{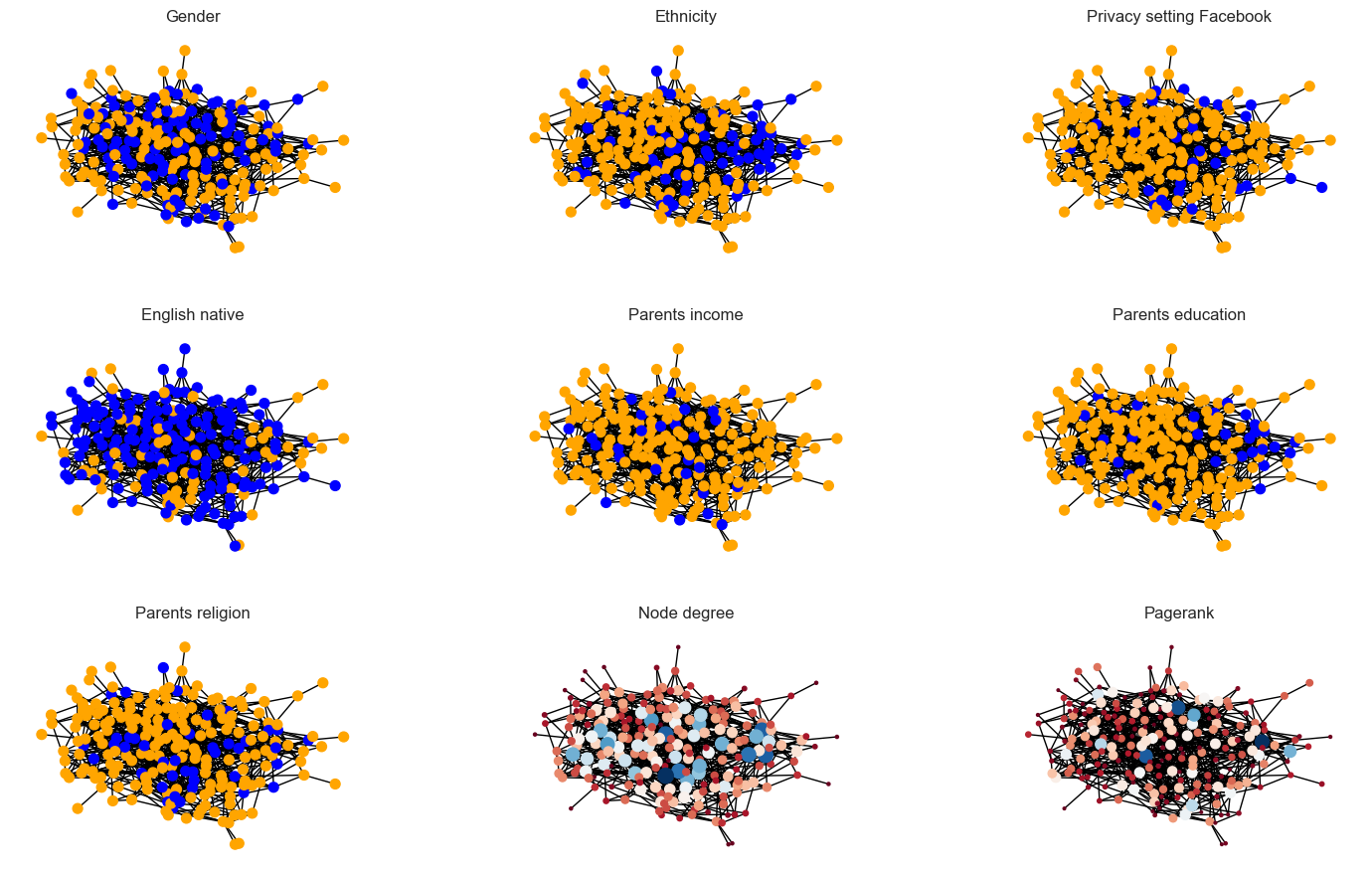}
    \caption{Representation of the student network with division into each of the minorities together with the node degree and pagerank value color map - blue indicating a higher value and red - a lower one.}
    \label{fig:minorities_networks}
\end{figure}

Similarly, Fig.\ref{fig:minorities_networks} presents a topological distribution of these minorities, together with a colormap of their node degrees and PageRank values.

\section{Opinion distribution for the worldview questions}
\label{appendix:sankey}
\begin{figure}[!htbp]
    \centering
    \includegraphics[width=0.95\linewidth]{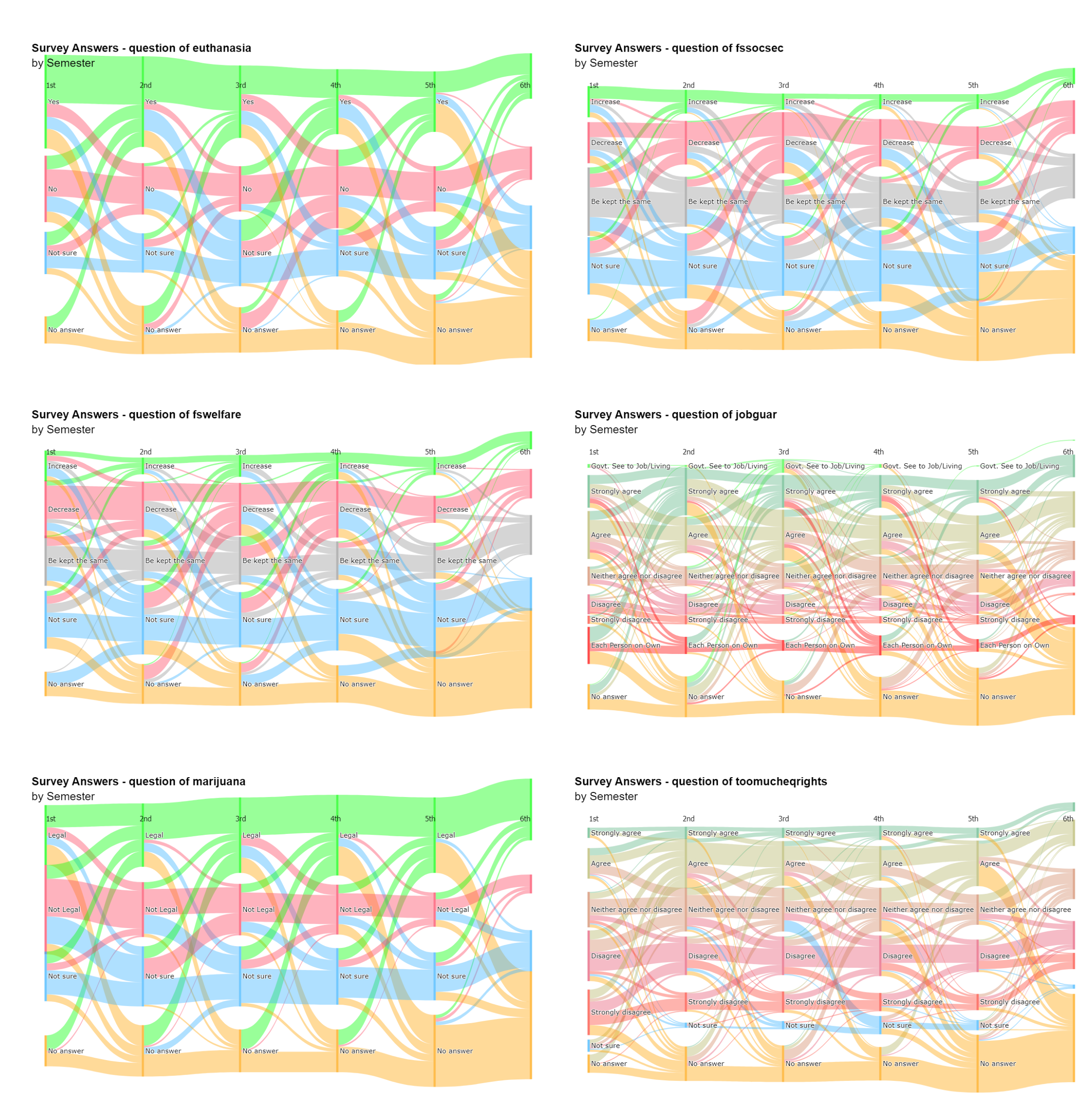}
    \caption{Diagram of the opinion change dynamics illustrating each of the six survey questions.}
    \label{fig:sankey_euthanasia}
\end{figure}
\par This Sankey graph illustrates how students' opinions have evolved over the span of 6 academic semesters for each of the six questions. Worth noting is the gradual increase in the yellow share of each of the answers - indicating people who have left the NetSense experiment before its planned end and did not fill later portion of the surveys.

\section{Overview of the Classifier Model Hyperparameters}
\label{appendix:hyperparameters}
\paragraph{Stratified Random Forest Hyperparameter Grid}
The grid for the Stratified Random Forest included:
\begin{itemize}
    \item \texttt{n\_estimators}: Number of trees in the forest, ranging from 50 to 400.
    \item \texttt{max\_features}: Number of features considered for splitting at each node, with options \texttt{\lq sqrt\rq} and \texttt{\lq log2'}.
    \item \texttt{max\_depth}: Maximum depth of trees, tested between 2 and 8.
    \item \texttt{min\_samples\_split}: Minimum number of samples required to split an internal node, ranging from 4 to 8.
    \item \texttt{min\_samples\_leaf}: Minimum number of samples required to be at a leaf node, ranging from 2 to 4.
    \item \texttt{bootstrap}: Boolean flag indicating whether to use bootstrapped samples (\texttt{True}) or the entire dataset (\texttt{False}).
    \item \texttt{criterion}: Splitting criterion, either \texttt{\lq gini'} or \texttt{\lq entropy'}.
\end{itemize}
\paragraph{Decision Tree Hyperparameters Grid}
The grid for the Decision Tree classifier, used for topology-based features, included:
\begin{itemize}
    \item \texttt{criterion}: Splitting criterion, with options \texttt{\lq gini'}, \texttt{\lq entropy'}, and \texttt{\lq log\_loss'}.
    \item \texttt{splitter}: Strategy used to choose the split at each node, with options \texttt{\lq best'} and \texttt{\lq random'}.
    \item \texttt{max\_depth}: Maximum depth of the tree, tested at values \texttt{1}, \texttt{25}, and \texttt{50}.
    \item \texttt{max\_features}: Number of features considered for splitting at each node, with options \texttt{\lq sqrt'}, \texttt{\lq log2'}, and \texttt{None}.
    \item \texttt{min\_samples\_leaf}: Minimum number of samples required to be at a leaf node, tested at values \texttt{1}, \texttt{5}, and \texttt{10}.
\end{itemize}


\bibliographystyle{elsarticle-harv} 
\bibliography{bibliography}

\end{document}